\documentstyle[aps,psfig,floats,twocolumn]{revtex}

\tighten

\def\fdag#1{#1\!\!\!/}

\begin{document}

\preprint{}
\draft

 \wideabs{      

\title{Preserving the gauge invariance of meson production currents\\ 
         in the presence of explicit final-state interactions}

\author{Helmut Haberzettl}

\address{Center for Nuclear Studies, Department of Physics, 
         The George Washington University, Washington, D.C. 20052}

\date{27 March 2000}

\maketitle

\begin{abstract}
  A comprehensive formalism is developed to preserve the gauge invariance of currents describing the
  photo- or electroproduction of mesons off the nucleon
  when the final-state interactions of mesons and nucleons is taken into account explicitly. 
  Replacing exchange currents by auxiliary currents,
  it is found that all contributions due to explicit final-state interactions are purely
  transverse and do not contain a Kroll--Ruderman-type contact current.
  The relation of the present formulation to tree-level-type prescriptions is shown.
\end{abstract}
\pacs{PACS numbers: 25.20.Lj, 24.10.Jv, 13.75.Gx, 24.10.Eq \hfill {\tiny [PRC{\bf 62},034605(2000)---nucl-th/0003058]}}

 }    

\section{Introduction}

\begin{figure}[b!]
\centerline{\psfig{file=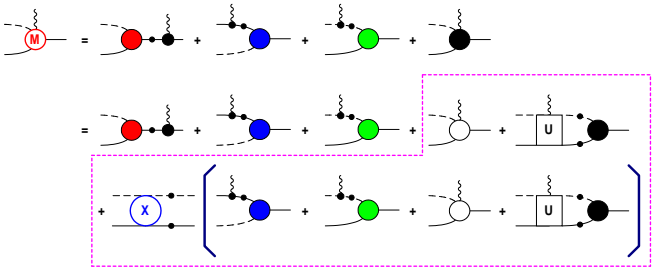,width=\columnwidth,clip=,silent=}}
\vspace{3mm}
\caption[Meson production diagrams]{\label{mfinal} 
    Meson production current $M^\mu$.
    Time proceeds from right to left.  The first line, which sums up the $s$-, $u$-, and $t$-channel diagrams
    and the interaction current $M^\mu_{\rm int}$ (right-most diagram), is referred to as the tree-level.
    The dynamical content of $M^\mu_{\rm int}$, including the final-state interaction mediated by the nonpolar 
    $\pi N$ amplitude $X$ (see Fig. \ref{pinndiagrams}),  is explicitly shown by the diagrams enclosed in
    the dashed box. The diagram element labelled $U$ subsumes all exchange currents $U^\mu$ contributing 
    to the process.  The diagram with open circle depicts the bare current $m^\mu_{\rm bare}$.}
\end{figure}

Gauge invariance is one of the central issues when attempting to describe how photons interact with hadronic systems.
Concentrating on the simplest case---pion photoproduction with real or virtual photons,
gauge invariance can easily be shown to follow if the
$\pi N$ and $\gamma N$ problems are treated completely and consistently on an equal footing 
\cite{WTI,kazes,antwerpen95,hh97g}.
In practice, however, one often needs to revert to some approximate treatment of one or more of the
contributing reaction mechanisms and this usually leads to a violation of gauge invariance.
To restore it, the neglected reaction mechanisms must be approximated by 
auxiliary currents constructed such that the gauge-invariance-violating contributions
to the four-divergence of the total production amplitude are cancelled. 
Such a procedure cannot be unique, of course, since one may always add arbitrary
transverse currents without affecting the four-divergence.

At the tree-level, where one does not resolve the internal mechanisms entering the interaction current 
(cf.\ Fig.\ \ref{mfinal}), various recipes exist to preserve gauge invariance.
The simplest case concerns the choice of bare vertices with pseudovector coupling for the $\pi NN$ vertex,
where the corresponding Kroll--Ruderman contact current \cite{krollruderman54} follows from
the minimal substitution procedure. The case of extended nucleons, whose internal structure is described in terms of
(phenomenological) form factors, is treated in Refs.\ \cite{hh97g,grossriska87,ohta89,nagorny,hh98tree}.

In the present work, we want to go beyond the tree level and investigate how one can preserve gauge invariance if
the internal structure of the interaction current is taken into account explicitly. The reaction mechanisms
that enter the interaction current are summarized within the dashed box in Fig.\ \ref{mfinal}. 
Specifically, we are interested in preserving gauge invariance in the explicit presence of 
hadronic final-state interactions. This is achieved by introducing auxiliary currents which
cancel the gauge-invariance-violating contributions. In particular,
we show how one may exploit the constraints following from the generalized Ward--Takahashi identities
to construct these currents. 

The present discussion is restricted to nucleons and pions only to facilitate the presentation.
We do not include here possible resonances or other transition mechanisms since
their couplings to the electromagnetic field are transverse and have no bearing
on the question of gauge invariance. None of these restrictions are essential, however,
and one may easily adapt the present formalism to accommodate more complex situations.

\section{Gauge Invariance}

\begin{figure}[t!]
\centerline{\psfig{file=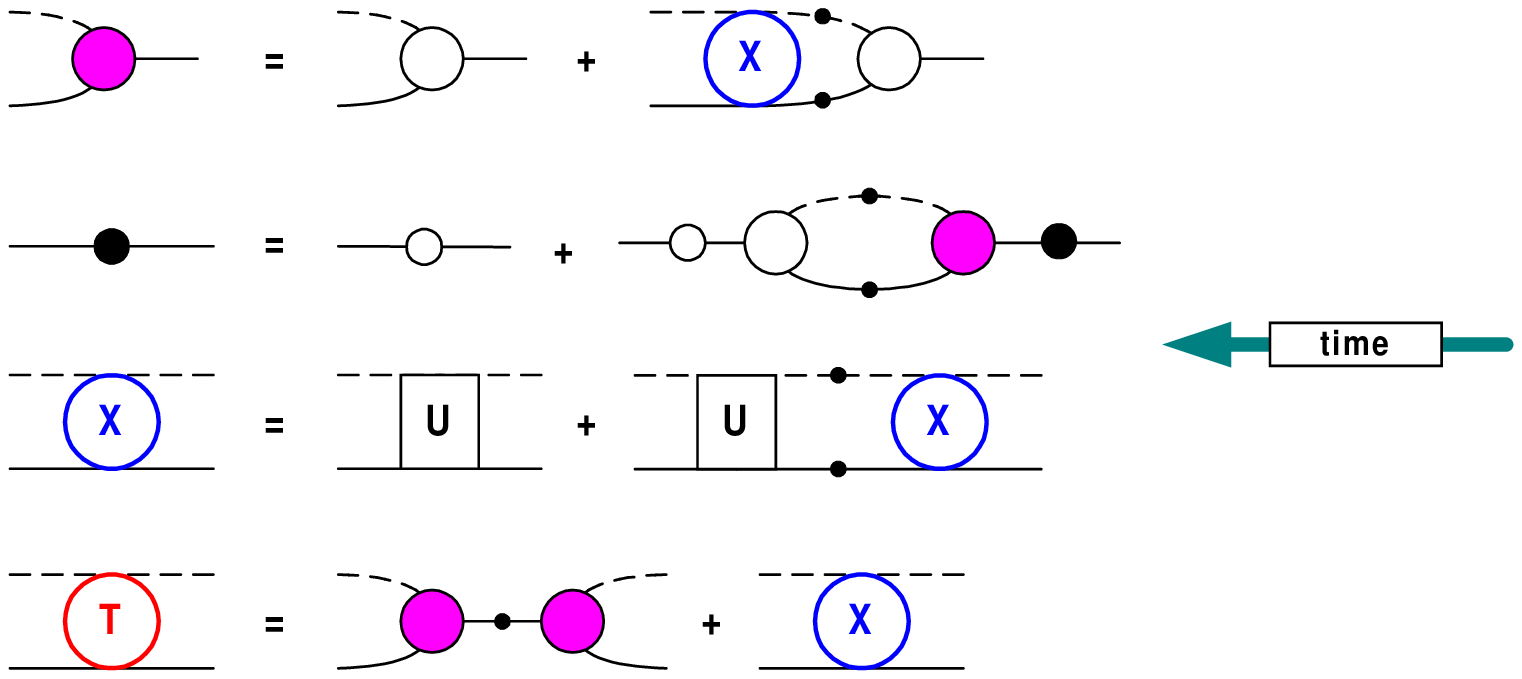,width=.9\columnwidth,clip=,silent=}}
\vspace{3mm}
\caption[Dressing diagrams]{\label{pinndiagrams} 
    Pion-nucleon scattering with fully dressed hadrons. The full $\pi N$-amplitude is 
    denoted by $T$, with $X$ subsuming all of its nonpolar (i.e., non-$s$-channel) contributions. 
    The latter satisfies the Bethe--Salpeter-type integral equation depicted in the third line here, where the
    driving term $U$ sums up all nonpolar irreducible contributions to $\pi N$-scattering, 
    i.e., all irreducible  contributions which do not contain an $s$-channel pole (see Ref.\ \cite{hh97g}
    for full details). Diagram elements with open, unlabeled circles describe bare quantities.}
\end{figure}

The pion photoproduction current of the nucleon is shown in Fig. \ref{mfinal} \cite{hh97g}.
According to the diagrams in the first line of this figure,
the total current $M^\mu$ may be broken up into four main contributions:
The three Born terms due to the $s$-, $u$-, and $t$-channel currents stemming 
from the photon coupling to the three external legs of the
$\pi NN$ vertex, and the interaction current $M^\mu_{\rm int}$ where the photon 
attaches itself to an internal leg of the $\pi NN$ vertex, i.e., 
\begin{equation}
M^\mu = M^\mu_s + M^\mu_u + M^\mu_t + M^\mu_{\rm int} \;.\label{M}
\end{equation}
While the first three contributions are relatively straightforward, the last one---as it is shown in
the last two lines of Fig.\ \ref{mfinal}---explicitly involves the full complexity
of the internal reaction dynamics of the underlying $\pi N$ scattering problem summarized in Fig.~\ref{pinndiagrams}.

For gauge invariance of the total current $M^\mu$ to hold true, 
its four-divergence must satisfy a generalized Ward--Takahashi identity \cite{kazes,antwerpen95,hh97g} 
\begin{eqnarray}
k_\mu M^\mu &=&  -[F_s\tau] S_{p+k}Q_i S^{-1}_p
 +S^{-1}_{p'}Q_f S_{p'-k} [F_u\tau] \nonumber\\
& &{ } +\Delta^{-1}_{q}Q_\pi\Delta_{q-k}[F_t\tau]  \;,\label{M_WTI}
\end{eqnarray}
where $p$ and $k$ are the four-momenta of the incoming nucleon and photon, respectively, and $p'$ 
and $q$ are the four-momenta of the outgoing nucleon and pion, respectively, 
related by momentum conservation $p'+q=p+k$. $S$ and $\Delta$ are the propagators of the
nucleons and pions, respectively, with their subscripts denoting the available four-momentum for the corresponding hadron;  
$Q_i$, $Q_f$, and $Q_\pi$ are the initial and final nucleon and the pion charge operators, respectively.
$[F_x\tau]$ denotes the $\pi NN$ vertex (including coupling and isospin operators), with the subscript $x$ labeling 
the kinematic situation appropriate for the $s$-, $u$-, or $t$-channel diagrams
appearing in Fig.\ \ref{mfinal}. The vertex isospin $\tau$
does not commute with the charge operators and
$\tau Q_i -Q_f \tau - Q_\pi \tau=0 $
describes charge conservation at the vertex in a symbolic manner.

Equation (\ref{M_WTI}) is easily obtained from Eq.\ (\ref{M}) upon using the Ward--Takahashi identities \cite{WTI} for
the nucleon and pion currents,
\begin{mathletters}\label{WTI}
\begin{eqnarray}
k_\mu J^\mu_{\sc n} &=& S^{-1}_{p_{\sc n}+k}Q_{\sc n} - Q_{\sc n} S^{-1}_{p_{\sc n}}\;,\\
k_\mu J^\mu_{\pi}   &=& \Delta^{-1}_{q_{\pi}+k}Q_{\pi} - Q_{\pi} \Delta^{-1}_{q_{\pi}}\;,
\end{eqnarray}
\end{mathletters}
where $p_{\sc n}$ and $q_\pi$ are the respective initial hadron momenta of the electromagnetic
vertices, and using the fact 
that for gauge invariance to be true the interaction current must obey \cite{kazes,hh97g} 
\begin{equation}
k_\mu M^\mu_{\rm int} = -[F_s\tau] Q_i  +Q_f[F_u\tau]+Q_\pi [F_t\tau] \;. \label{WTI_int}
\end{equation}

\subsection{Preserving gauge invariance}

As alluded to above, the preceding relations can easily be shown to be true if the 
$\pi N$ and $\gamma N$ problems are treated consistently on an equal footing \cite{antwerpen95,hh97g}.
In practical applications, however, approximations are inevitable which usually violate
the gauge invariance.

To see how one may preserve gauge invariance in such a situation, let us
explicitly write the four-divergence of the interaction current using
the relevant parts of Fig.~\ref{mfinal} as guidance.
One finds
\begin{eqnarray}
k_\mu M^\mu_{\rm int} &=& k_\mu m^\mu_{\rm bare} +k_\mu U^\mu G_0 [F \tau]\nonumber\\
& &{ }+ XG_0 \Big\{ 
    k_\mu J^\mu_{\sc n} S_{p_{\sc n}-k}   [F_u \tau]  
   +k_\mu J^\mu_\pi \Delta_{q_\pi-k}  [F_t \tau] \nonumber\\
& &{ }\quad\quad      +k_\mu m^\mu_{\rm bare} +k_\mu U^\mu G_0 [F \tau]      \Big\} \;, \label{WTI_int_expl}
\end{eqnarray}
where $G_0=S_{p_{\sc n}}\Delta_{q_\pi}$ denotes the product of the intermediate nucleon and pion propagators, with respective
four-momenta $p_{\sc n}$ and $q_\pi$ denoting the integration variables, and $X$ is the nonpolar
$\pi N$ amplitude (see Fig.~\ref{pinndiagrams}) which mediates the hadronic final-state interaction;
$U^\mu$ subsumes all exchange currents and $m^\mu_{\rm bare}$ is the bare contact
current. 

For gauge invariance to hold true,
the bare current must satisfy the condition \cite{hh97g}
\begin{eqnarray}
k_\mu m^\mu_{\rm bare} = -[f_s\tau]Q_i + Q_f [f_u\tau] + Q_\pi [f_t\tau] \;,\label{WTI_bare}
\end{eqnarray}
where $[f_x \tau]$ denotes the bare $\pi NN$ vertex the same way the notation $[F_x \tau]$
was used above for the dressed vertex. This is the analog of Eq.\ (\ref{WTI_int})
for the bare current; it is usually satisfied as a matter of course. One of the simplest 
nontrivial examples is the case of pure pseudovector coupling
without form factors, where $m^\mu_{\rm bare}$ is the Kroll--Ruderman contact
current \cite{krollruderman54};  see Eq.\ (\ref{kr}). 

Combining now Eq.\ (\ref{WTI_int_expl}) with the necessary condition
(\ref{WTI_int}), and making use of Eq.\ (\ref{WTI_bare}) in the Born terms, produces
\begin{eqnarray}
0 &=& \Big([F_s\tau]-[f_s\tau]\Big) Q_i - Q_f \Big([F_u\tau]-f_u\tau]\Big)  \nonumber\\
& &{ }- Q_\pi\Big([F_t\tau] -[f_t\tau] \Big) +k_\mu U^\mu G_0 [F \tau]\nonumber\\
& &{ }+ XG_0 \Big\{
    k_\mu J^\mu_{\sc n} S_{p_{\sc n}-k}   [F_u \tau]  
   +k_\mu J^\mu_\pi \Delta_{q_\pi-k}  [F_t \tau]
           \nonumber\\
& &{ }  \quad\quad+ k_\mu m^\mu_{\rm bare}+k_\mu U^\mu G_0 [F \tau]      \Big\}  \label{WTI_condition}
\end{eqnarray}
as a necessary {\it off-shell} requirement for gauge invariance to be satisfied.

In other words, 
as long as the basic Ward--Takahashi identities (\ref{WTI}) for the hadron
currents are true,
{\it any} approximation of the full reaction mechanisms constructed in such a manner that the
condition (\ref{WTI_condition}) is satisfied will also preserve gauge invariance as a matter of course.

In view of the arbitrariness of transverse contributions, 
there are of course infinitely many ways this can be achieved. The prescription we
give in the following applies to the simplifying assumption that one 
completely omits the explicit treatment of exchange currents $U^\mu$.  
However, even if they are taken into account in some partial manner, 
it is a straightforward exercise to adapt the following 
formulation to accommodate such situations.

Omitting explicit exchange currents,
we maintain gauge invariance by constructing auxiliary currents $j^\mu_0$
and $j^\mu_1$ which provide the same effect as the exchange currents $U^\mu$ 
as far as the preservation of gauge invariance is concerned. 

To this end, we make the replacements
\begin{mathletters}
\begin{eqnarray}
U^\mu G_0 [F \tau] &\longrightarrow& j^\mu_0 +\Delta j^\mu_0\;,\\
XG_0 U^\mu G_0 [F \tau] &\longrightarrow& j^\mu_1+\Delta j^\mu_1\;,
\end{eqnarray}
\end{mathletters}
and demand that $j^\mu_0$ and $j^\mu_1$ satisfy
\begin{eqnarray}
k_\mu j^\mu_0 &=& -\Big([F_s\tau]-[f_s\tau]\Big) Q_i 
                  + Q_f \Big([F_u\tau] -[f_u\tau] \Big) \nonumber\\
& &{ }+ Q_\pi\Big([F_t\tau] -[f_t\tau] \Big) \label{split0}
\end{eqnarray}
and
\begin{eqnarray}
k_\mu j^\mu_1  &=&- XG_0 \Big\{ 
           k_\mu J^\mu_{\sc n} S_{p_{\sc n}-k}   [F_u \tau]  
   +k_\mu J^\mu_\pi \Delta_{q_\pi-k}  [F_t \tau]\nonumber\\
& &{ }\quad\quad      +k_\mu m^\mu_{\rm bare}   \Big\}    \;. \label{split1}
\end{eqnarray}
Clearly, if these conditions are met, then 
\begin{equation}
k_\mu \left(\Delta j^\mu_0+\Delta j^\mu_1\right) = 0\;, \label{delta_transverse}
\end{equation}
i.e., $\Delta j^\mu_0+\Delta j^\mu_1$ is purely transverse. 
Without explicit treatment of exchange currents, these contributions
are inaccessible and will be dropped. The resulting production current,
\begin{eqnarray}
M^\mu &=& M^\mu_s + M^\mu_u + M^\mu_t + m^\mu_{\rm bare} +j^\mu_0 \nonumber\\
    & &{ } + XG_0 \Big\{M^\mu_u + M^\mu_t + m^\mu_{\rm bare}  \Big\} +j^\mu_1 \;, \label{M_approx}
\end{eqnarray}
will then satisfy the generalized Ward--Takahashi identity (\ref{M_WTI}) 
and therefore it will be gauge invariant.

To be more specific as to how to implement the conditions
 (\ref{split0}) and (\ref{split1}), allowing for a mixture of pseudoscalar and pseudovector couplings,
let us write the dressed $\pi NN$ vertex as
\begin{equation}
F = g_{\rm ps} \gamma_5 G_{\rm ps} + g_{\rm pv} \frac{\gamma_5 \fdag{q}_\pi}{2m} G_{\rm pv}\;,
\end{equation}
where the indices ps and pv stand for pseudoscalar and pseudovector contributions, respectively;
$G_{\rm ps}$ and $G_{\rm pv}$ denote the corresponding normalized form factors 
(with their strength parameters  $g_{\rm ps}$ and $g_{\rm pv}$ adding up to the
physical coupling constant, $g_{\pi NN}=g_{\rm ps}+g_{\rm pv}$),
$q_\pi$ is the four-momentum of the pion, and $m$ the nucleon mass.
The bare vertex $f$ is given by the same equation  with all $G$'s removed, and
the corresponding bare current,
\begin{equation}
m^\mu_{\rm bare} = -g_{\rm pv} \frac{\gamma_5 \gamma^\mu}{2m} Q_\pi \tau\;, \label{kr}
\end{equation}
 is just the usual Kroll--Ruderman contact term \cite{krollruderman54}.

Equation (\ref{split0}) now reads explicitly
\begin{eqnarray}
k_\mu j^\mu_0 &=& -\gamma_5\Big(G_s - g \Big) \tau Q_i \nonumber\\
& &{ }           + \gamma_5\Big(G_u - g \Big)Q_f \tau \nonumber\\
& &{ }           + \gamma_5\Big(G_t - g \Big) Q_\pi \tau \nonumber\\
& &{ }           - g_{\rm pv} \gamma_5 \frac{\fdag{k}}{2m}\Big(G_{{\rm pv},s}\tau Q_i-G_{{\rm pv},u}Q_f\tau \Big)\nonumber\\[2mm]
& &{ }           - k_\mu m^\mu_{\rm bare} \;,\label{step1_split0}
\end{eqnarray}
where 
\begin{equation}
G_x = g_{\rm ps} G_{{\rm ps}, x}+g_{\rm pv} \frac{\fdag{p}-\fdag{p}'}{2m} G_{{\rm pv},x} \label{G_x}
\end{equation}
(with $x=s$, $u$, $t$) denotes the kinematic situations in which the 
vertex functions $G_{\rm ps}$ and $G_{\rm pv}$ appear; 
$g$ is given by the same equation with all $G$'s removed.

Note that all the terms containing $g$ in Eq.\ (\ref{step1_split0}) add up to zero. We may therefore
replace $g$ by an arbitrary function $\widehat{F}$ with impunity. Furthermore,
using the Mandelstam variables
$s=(p+k)^2$, $u=(p'-k)^2$, and $t=(q-k)^2$,
we may then rewrite the resulting equation as
\begin{eqnarray}
k_\mu j^\mu_0 &=& k_\mu \Bigg\{ -\frac{(2p+k)^\mu}{s-p^2}\gamma_5\Big(G_s - \widehat{F} \Big) \tau Q_i   \nonumber\\
& &{ }\quad\quad             - \frac{(2p'-k)^\mu}{u-p'^2}\gamma_5\Big(G_u - \widehat{F} \Big) Q_f \tau   \nonumber\\[2mm]
& &{ }\quad\quad               - \frac{(2q-k)^\mu}{t-q^2}\gamma_5\Big(G_t - \widehat{F} \Big) Q_\pi \tau \nonumber\\
& &{ }\quad\quad           - g_{\rm pv} \frac{\gamma_5 \gamma^\mu}{2m}\Big(G_{{\rm pv},s}\tau Q_i-G_{{\rm pv},u}Q_f\tau\Big)\nonumber\\
& &{ }\quad\quad           - m^\mu_{\rm bare}\Bigg\} \;,\label{step2_split0}
\end{eqnarray}
which allows us to put
\begin{eqnarray}
j^\mu_0 &=&  -\frac{(2p+k)^\mu}{s-p^2}\gamma_5\Big(G_s - \widehat{F} \Big) \tau Q_i \nonumber\\
& &{ }           - \frac{(2p'-k)^\mu}{u-p'^2}\gamma_5\Big(G_u - \widehat{F} \Big)Q_f \tau \nonumber\\
& &{ }           - \frac{(2q-k)^\mu}{t-q^2}\gamma_5\Big(G_t - \widehat{F} \Big) Q_\pi \tau \nonumber\\
& &{ }   -  g_{\rm pv} \frac{\gamma_5 \gamma^\mu}{2m}\Big(G_{{\rm pv},s}\tau Q_i-G_{{\rm pv},u}Q_f\tau\Big)
               - m^\mu_{\rm bare} \;.\label{step3_split0}
\end{eqnarray}
Comparing with Eq.\ (\ref{M_approx}), note that the last two terms of this gauge-invariance-preserving current cancel the
bare Kroll--Ruderman term and replace it by a dressed one, where instead of the pion charge $Q_\pi\tau$,
there is now a dressing term $G_{{\rm pv},s}\tau Q_i-G_{{\rm pv},u}Q_f\tau$. This dressing term expresses
the pion charge in terms of the nucleon charges modified by hadronic form factors, and, in general, it will be non-zero even if the
pion is uncharged.

We emphasize that the transition from Eq.\ (\ref{step1_split0}) to (\ref{step3_split0}) is not unique, of course,
since one may add a divergence-free current to $j^\mu_0$ without changing the necessary 
condition (\ref{step1_split0}). In fact, the replacement of $g$ by $\widehat{F}$ amounts to
the addition of such a transverse current (which in turn may be understood
as a phenomenological way of getting a handle on the neglected transverse current $\Delta j^\mu_0+\Delta j^\mu_1$). 
At this stage, $\widehat{F}$ is completely undetermined.
Below, when considering the relationship of the present results to existing tree-level
approaches, we will discuss some specific choices.


We also note that the terms appearing in Eq.\ (\ref{step3_split0}) do not introduce any new singularities
into the amplitude. This is easily illustrated for the example of the $s$-channel pole diagram
$M^\mu_s$, {\it viz.}
\begin{eqnarray}
M^\mu_s&=& [F_s\tau] \frac{\fdag{p}+\fdag{k}+m}{s-m^2} \gamma^\mu  Q_i
  + M^\mu_{{\sc t},s}
\nonumber\\
&=& \gamma_5 G_s \frac{(2p+k)^\mu}{s-m^2}   \tau Q_i
   + g_{\rm pv} \frac{\gamma_5\fdag{k}}{2m} G_{{\rm pv},s} 
   \frac{(2p+k)^\mu}{s-m^2}   \tau Q_i
\nonumber\\
& &{ }
+\gamma_5 G_{s,\fdag{q}}
   \frac{k^\mu-\gamma^\mu \fdag{k}}{s-m^2}  \tau Q_i
  + M^\mu_{{\sc t},s}\;,
\end{eqnarray}
where $G_{s,\fdag{q}}$ is given by Eq.\ (\ref{G_x}) with $\fdag{p}-\fdag{p}'$ replaced 
by $\fdag{q}$; $M^\mu_{{\sc t},s}$ splits off the transverse
part of the electromagnetic nucleon current given in Eq.\ (\ref{transverse_N}) below.
The decomposition given here makes it immediately obvious that the effect of adding
the $s$-channel term,  with $p^2=m^2$, of Eq.\ (\ref{step3_split0}) to this expression is to replace  $G_s$ in the
first term here by $\widehat{F}$. 
The same happens also for the $u$- and $t$-channel diagrams. In other words, apart
from providing a dressed Kroll--Ruderman term, the effect of the gauge-invariance preserving
current  (\ref{step3_split0}) is to provide a common form factor $\widehat{F}$ for some
(but not all) of the contributions originally containing individual form factors $G_s$, $G_u$, and $G_t$,
without changing the original singularities of the Born diagrams.


Next, to construct the current $j^\mu_1$, we recall that
the gauge-invariant nucleon and pion currents appearing in the $u$- and $t$-channel terms
of the final-state interaction contribution are given by
\begin{mathletters}\label{currents}
\begin{eqnarray}
J^\mu_{\sc n} &=& \gamma^\mu Q_{\sc n}+J^\mu_{{\sc t},{\sc n}}     \;,  \\
J^\mu_\pi &=& (2q_\pi-k)^\mu Q_\pi + J^\mu_{{\sc t},\pi}  \;,
\end{eqnarray}
\end{mathletters}
respectively, with transverse pieces which follow from demanding the validity of the
Ward--Takahashi identities (\ref{WTI}), i.e.,
\begin{mathletters}\label{transverse}
\begin{eqnarray}
J^\mu_{{\sc t},{\sc n}} &=&  \left(\gamma^\mu-k^\mu \frac{\fdag{k}}{k^2} \right)Q_{\sc n}(F_1-1) +i\frac{\sigma^{\mu\nu}k_\nu}{2m}\kappa F_2\;,
              \label{transverse_N}       \\
J^\mu_{{\sc t},\pi} &=& \left[(2q_\pi-k)^\mu -k^\mu \frac{k \cdot (2q_\pi-k)}{k^2}\right]Q_\pi (F_\pi-1)\;,
                      \label{transverse_pi}\nonumber\\
\end{eqnarray}
\end{mathletters}
where $F_1$ and $F_2$ respectively are the electromagnetic Dirac and Pauli form factors of the nucleon,
with $\kappa$ being its anomalous magnetic moment, and $F_\pi$ is the electromagnetic form factor of the pion.
Their four-divergence is given by
\begin{mathletters}\label{WTI_diff}
\begin{eqnarray}
k_\mu J^\mu_{\sc n} &=&  k_\mu \gamma^\mu Q_{\sc n}\;,\\
k_\mu J^\mu_\pi     &=&  k_\mu (2q_\pi-k)^\mu Q_\pi\;.
\end{eqnarray}
\end{mathletters}
Inserting this into Eq.\ (\ref{split1}), we may then extract the
gauge-invariance-preserving current as
\begin{eqnarray}
 j^\mu_1 
&=& - X G_0\Big\{\gamma^\mu  Q_{\sc n}   S_{p_{\sc n}-k}[F_u\tau]\nonumber\\
& &{ }\quad\quad  + (2q_\pi-k)^\mu  Q_\pi \Delta_{q_\pi-k} [F_t\tau]        
+ m^\mu_{\rm bare}  \Big\} \;. \label{j1_GI}
\end{eqnarray}
Its effect on the final-state interaction part of the
production amplitude (\ref{M_approx}) is seen to
simply cancel the bare current
and to reduce the $u$- and $t$-channel contributions 
to their respective transverse pieces%
, i.e.,
\begin{eqnarray}
XG_0 \Big(M^\mu_{{\sc t},u} + M^\mu_{{\sc t},t}\Big) 
&=& XG_0 \Big\{M^\mu_u + M^\mu_t + m^\mu_{\rm bare}  \Big\} +j^\mu_1 \nonumber\\
&=&{ } XG_0 \Big\{
    J^\mu_{{\sc t},{\sc n}} S_{p_{\sc n}-k}   [F_u \tau]  \nonumber\\
& &{ }\quad\quad   +J^\mu_{{\sc t},\pi} \Delta_{q_\pi-k}  [F_t \tau]  \Big\}\;.
\end{eqnarray}
The entire contribution from the final-state interaction, therefore, is purely transverse.

\subsection{Relation to Ohta's and Haberzettl's tree-level prescriptions}

To make the connection with tree-level approaches, we need to switch off all final-state interactions
and put $X=0$ in Eqs.\ (\ref{split1}) and (\ref{M_approx}). 
The conditions to be satisfied, therefore, are Eq.\ (\ref{split0}) and Eq.\ (\ref{split1})
in the form
\begin{equation}
k_\mu j^\mu_1 = 0 \;. \label{split1_tree}
\end{equation}
In other words, simply putting $j^\mu_1=0$ satisfies all gauge-invariance constraints at the tree level.  

Both Ohta's \cite{ohta89} and Haberzettl's \cite{hh97g,hh98tree} prescriptions for
preserving gauge invariance can be understood as different choices for the function $\widehat{F}$
in Eq.\ (\ref{step3_split0}). 

Ohta's approach, based on a particular application of the minimal substitution procedure, finds
\begin{equation}
\widehat{F} =  g_{\rm ps} G_{\rm ps}(q,p',p)+g_{\rm pv} \frac{\fdag{p}-\fdag{p}'}{2m} G_{\rm pv}(q,p',p)\;, \label{ohta}
\end{equation}
where the external hadron momenta of the photoproduction current---a four-point function---appear here
as the momenta of the $\pi NN$ vertex---a three-point function. Since the momenta satisfy 
$q+p'=p+k$, this mismatch corresponds to an unphysical region of the $\pi NN$ vertex (which leads to the problems
discussed in Ref.\ \cite{wang96}). Only in the infrared limit of $k=0$, this mismatch is resolved and then
this choice prevents the current (\ref{step3_split0}) from being singular at $k=0$.

In Haberzettl's prescription, the function $\widehat{F}$ is a linear combination of the
three kinematical situations in which the $\pi NN$ vertices appear in the Born terms, i.e., 
[cf.\ Eq. (\ref{G_x})]
\begin{equation}
\widehat{F} = a_s G_s + a_u G_u + a_t G_t \;, \label{hh}
\end{equation}
with coefficients constrained by $a_s + a_u + a_t=1$, which may be fixed according to
prejudice or used as free fit parameters. In contrast to Ohta's choice, this does not
require any unphysical values for the $\pi NN$ form factors in practical applications,
and it also has a well-behaved infrared limit.
In direct comparisons to Ohta's, this prescription
is found to provide better agreement with the experimental data
\cite{hh98tree,feuster99,han99} .

\section{Summary and Discussion}

In summary, we have treated here the electromagnetic production current for mesons off the nucleon
for both real and virtual photons. We used the constraints following from requiring the
validity of the generalized Ward--Takahashi identities to
construct auxiliary current pieces that ensure that gauge invariance is preserved
even if---as it is invariably the case in practical applications---one does not treat the problem completely and consistently.
The result for the production current $M^\mu$ obtained here 
in Eqs.\ (\ref{M_approx}), (\ref{step3_split0}), and (\ref{j1_GI})
can be summarized by 
\begin{eqnarray}
M^\mu &=& M^\mu_s + M^\mu_u + M^\mu_t  +j^\mu_{\sc gip}   \nonumber\\
      & &{ } + XG_0 \Big(M^\mu_{{\sc t},u} + M^\mu_{{\sc t},t}  \Big) \;, \label{M_GI}
\end{eqnarray}
where the gauge-invariance-preserving current,
\begin{equation}
j^\mu_{\sc gip} = j^\mu_0 + m^\mu_{\rm bare}  \;,
\end{equation}
is given via $j^\mu_0$ of  Eq.\ (\ref{step3_split0}). 

As far as the choice of the function $\widehat{F}$ appearing in $j^\mu_0$ is concerned,
the generally better results obtained with Haberzettl's prescription 
at the tree-level \cite{hh98tree,feuster99,han99} seem to favor the form given in Eq.\ (\ref{hh}).
We emphasize, however, that other functions are possible 
and that, in general, as far as gauge invariance is concerned, any current $j^\mu_0$ that satisfies
the necessary  condition (\ref{step1_split0}) is permitted here.

The part describing the hadronic final-state interaction due to $X$ 
does no longer contain the Kroll--Ruderman contact term since 
it is canceled by the current of Eq.\ (\ref{j1_GI}).
Moreover, what remains is seen to be entirely
transverse since the corresponding $u$- and $t$-channel contributions, 
$M^\mu_{{\sc t},u}$ and $M^\mu_{{\sc t},t}$, respectively contain only
the transverse electromagnetic nucleon and pion operators of
Eq.\ (\ref{transverse}). For real photons, in particular, 
both  $F_\pi-1$ and  $F_1-1$ vanish and therefore $M^\mu_{{\sc t},t}=0$, i.e., the $t$-channel
does not contribute at all, and the $u$-channel current $M^\mu_{{\sc t},u}$ is reduced to the magnetic
$\sigma^{\mu\nu}k_\nu$ term from  Eq. (\ref{transverse_N}).

We emphasize that even though the gauge-invariant production current of Eq.\ (\ref{M_GI}) is 
an approximation to the full dynamics of the problem as summarized in
Figs. \ref{mfinal} and \ref{pinndiagrams}, it is complete as far as the
longitudinal components of the current are concerned. Any additional pieces must be
transverse. 

Let us repeat once again that the present work was restricted to pions and nucleons
merely to simplify the presentation. The concepts developed here are quite general, however.
An extension to other baryons and mesons, therefore, is straightforward and
easily done along the lines given here.
In particular, the fact that the final-state interaction contributions are purely
transverse remains true if one takes into account additional intermediate hadrons
($\Delta$,  $\rho$, etc.) since, just as was demonstrated here for the nucleons and pions,
only the transverse parts of their respective current operators survive in the final-state
interaction terms.

Finally, let us point out that the present results remain equally valid whether the
form factors $F$ and the final-state amplitude $X$ are obtained via some
sophisticated Bethe--Salpeter-type formalism or are based on a simple
phenomenological model ansatz. How these elements are obtained does not enter
any of the present considerations and therefore has no bearing on the question
of gauge invariance.

\acknowledgments
The author gratefully acknowledges discussions with S. Krewald and 
K. Nakayama which precipitated the present work.
This work was supported in part by Grant No.\ DE-FG02-95ER-40907 of 
the U.S. Department of Energy.

\end{document}